\documentclass[aps,prl,superscriptaddress,long,notitlepage,balancelastpage,nofootinbib,floatfix,twocolumn]{revtex4-2}
\pdfoutput=1
\usepackage{amsmath,mathtools,amssymb,amsthm,amsxtra,overpic,bbm,epsfig,subfigure,url,bm}
\usepackage{hyperref}
\usepackage{mathrsfs}
\usepackage{color,xcolor}
\usepackage{comment}
\usepackage{float}
\usepackage{enumitem}
\usepackage{slashed}
\usepackage{multirow}
\usepackage{appendix}
\usepackage{amsmath}
\usepackage[customcolors]{hf-tikz}
\DeclareUnicodeCharacter{03BC}{\ensuremath{\mu}}
\DeclareUnicodeCharacter{2032}{\ensuremath{'}}
\DeclareUnicodeCharacter{2009}{\,}
\DeclareUnicodeCharacter{3B1}{\ensuremath{\alpha}}
\definecolor{nicered}{rgb}{0.5,0.,0.}
\definecolor{nicegreen}{rgb}{0.,0.5,0.}
\definecolor{niceblue}{rgb}{0.,0.,0.5}
\hypersetup{
	colorlinks=true,
	linkcolor=black,
	filecolor=nicegreen,      
	urlcolor=niceblue,
	citecolor=nicered,
}

\makeatletter
\newcommand*{\balancecolsandclearpage}{%
	\close@column@grid
	\cleardoublepage
	\twocolumngrid
}
\makeatother
\begin{document}
\title{\Large 
Restricting Sterile Neutrinos by \\
Neutrinoless Double Beta Decay
}
\author{\bf Sudip Jana}
\email[E-mail: ]{sudip.jana@mpi-hd.mpg.de}
\affiliation{Max-Planck-Institut f{\"u}r Kernphysik, Saupfercheckweg 1, 69117 Heidelberg, Germany}

\author{\bf Lucas Puetter}
\email[E-mail: ]{lucas.puetter@mpi-hd.mpg.de}
\affiliation{Max-Planck-Institut f{\"u}r Kernphysik, Saupfercheckweg 1, 69117 Heidelberg, Germany}

\author{\bf Alexei Yu. Smirnov}
\email[E-mail: ]{smirnov@mpi-hd.mpg.de}
\affiliation{Max-Planck-Institut f{\"u}r Kernphysik, Saupfercheckweg 1, 69117 Heidelberg, Germany}

\begin{abstract}
The bounds on parameters of the eV and higher scale sterile neutrinos from the \( 0\nu \beta \beta \) decay have been refined and updated. We present a simple and compact analytic expression for the bound in the $\Delta m^2_{41} - \sin^2 2\theta_{14}$ plane, which includes all relevant parameters. Dependencies of the bound on unknown CP-phases and the type of mass spectrum of light neutrinos (mass ordering and level of degeneracy) are studied in detail. We have computed the bounds using the latest and most stringent data from  \texttt{KamLAND-Zen}. The projected constraints from future experiments are estimated. The obtained bounds are confronted with positive indications of the presence of sterile neutrinos as well as with 
the other existing bounds. The \( 0\nu \beta \beta \) decay results exclude the regions of parameters implied by BEST and Neutrino-4 and the regions indicated by LSND and MiniBooNE are in conflict 
with $0\nu \beta \beta$ results combined with $\nu_\mu-$ disappearance bounds.
\noindent 
\end{abstract}
\maketitle
\textbf{\emph{Introduction}.--} 
Despite the strong skepticism and controversy regarding the existence of the eV scale sterile neutrinos, considerable interest in these neutrinos still remains. Prompted by LSND \cite{LSND:2001aii}, and MiniBooNE \cite{MiniBooNE:2020pnu}, and then by the Gallium  calibration anomaly \cite{GALLEX:1997lja, Kaether:2010ag, SAGE:1998fvr, Abdurashitov:2005tb, Acero:2007su, Giunti:2010zu, Kostensalo:2019vmv}, and some reactor antineutrino experiments \cite{Mention:2011rk}, the existence of sterile neutrinos was further supported by the Neutrino-4  \cite{NEUTRINO-4:2018huq, Serebrov:2020kmd}, and BEST data \cite{Barinov:2022wfh}. At the same time, these positive signals are in contradiction with a number of bounds from disappearance oscillation experiments at accelerators and reactors \cite{Danilov:2022bss, DayaBay:2024nip, Andriamirado:2024glk, RENO:2020hva, STEREO:2022nzk}, as well as experiments with solar neutrinos and KATRIN \cite{KATRIN:2022ith}. Moreover,  the cosmological observables, such as the CMB, BBN, and large-scale structures, exclude these neutrinos unless some new physics is invoked that suppresses their production in the early universe (see \cite{Abazajian:2012ys, Acero:2022wqg} for reviews). Nevertheless,  various forthcoming and future experiments are aimed to test the existence of these neutrinos. Furthermore, sterile neutrinos may play an important role in the theory of usual neutrino masses and mixing \cite{Balaji:2001ns, our}. 

Substantial restrictions on the mixing of sterile neutrinos with active neutrinos can be obtained from the $0\nu\beta\beta$ decay under the assumption that neutrinos are Majorana particles. Indeed, due to this mixing, the eV scale eigenstate contributes to the $0\nu\beta\beta$ decay, and therefore, the mixing and mass can be restricted by the $0\nu\beta\beta$ experimental data. A number of aspects of this restriction were discussed \cite{Benes:2005hn, Goswami:2005ng, Li:2011ss, Barry:2011wb,  Giunti:2012tn, Girardi:2013zra, Pascoli:2013fiz, Guzowski:2015saa, Giunti:2015kza, Jang:2018zug, Huang:2019qvq, Majhi:2019hdj}. This includes the upper bounds in the $\Delta m^2_{41} - \sin^2 2\theta_{14}$ plane \cite{Giunti:2012tn, Jang:2018zug}, cancellation of contributions of light and heavy neutrinos and the implications of a vanishing effective Majorana mass \cite{Li:2011ss, Barry:2011wb, Pascoli:2013fiz, Girardi:2013zra}, the effect of Majorana phases \cite{Goswami:2005ng}, the role of uncertainties in nuclear matrix elements \cite{Guzowski:2015saa}, and uncertainties in mixing parameters of active neutrinos \cite{Giunti:2015kza}.

In this paper, we refine and update the constraints on sterile neutrinos from $0\nu\beta\beta$ decay. We derive a simple and compact formula for the edges of excluded regions. We identify the relevant physical quantities and consider their determination. Dependencies of bounds on the CP-phases, mass ordering, and absolute mass scale of active neutrinos are studied. We use the latest bound on the effective Majorana mass from the KAMLAND-Zen experiment, which is at the level of contribution of light neutrinos extracted from oscillation data for both mass hierarchies. We discuss the implications of the bounds for the Neutrino-4, BEST, and Gallium anomalies. In addition, we apply the obtained bounds to the  LSND and MiniBooNE anomalies interpreted as the transitions $\stackrel{(-)}{\nu_\mu} \rightarrow \stackrel{(-)}{\nu_e}$. While our study focuses on eV scale sterile neutrinos, the results are also applicable to sterile neutrinos with masses up to the MeV scale. 
\vspace{0.1 in}

\textbf{\emph{Effective Majorana mass and sterile neutrino parameters}.-- }
We consider the  (3+1) Majorana neutrino scheme, in which three active neutrinos, \(\nu_\alpha = (\nu_e, \nu_\mu, \nu_\tau)^T\)  mix with a single sterile neutrino, \(\nu_S\).  The $ee-$element of the $4 \times 4$ Majorana mass matrix $m_{ee}^0$  can be expressed in terms of  masses and mixing  as
\begin{equation}
|m_{ee}^0| e^{i\gamma} =  \sum_{j = 1 - 4}  |U_{ej}|^2 m_j e^{i\alpha_j}, 
\label{eq:1}
\end{equation}
where $|U_{ej}|$ are the moduli of elements of the mixing matrix, $m_j$ are the moduli of mass eigenvalues, and $\alpha_j$ are the complex phases, which include the Dirac phases originating from mixing matrix elements and the Majorana phases associated to masses. Separating the contribution of the heaviest eigenstate, the expression (\ref{eq:1}) can be rewritten as 
\begin{equation}
|m_{ee}^0| e^{i\gamma} =  |m_{ee}| e^{i\phi} +  |U_{e4}|^2 m_4 e^{i\alpha_4}, 
\label{eq:mass4}
\end{equation}
where 
\begin{equation}
|m_{ee}| e^{i\phi} \equiv \sum_{i = 1-3}  |U_{ei}|^2 m_i e^{i\alpha_i}
\label{eq:mass3}
\end{equation}
is the contribution of three light mass eigenstates to $m_{ee}^0$, while $m_4$ and $U_{e4}$ are the mass and mixing of the eV scale neutrino. 

\begin{figure*}[htb!]
\centering
  \includegraphics[width=0.90 \textwidth]{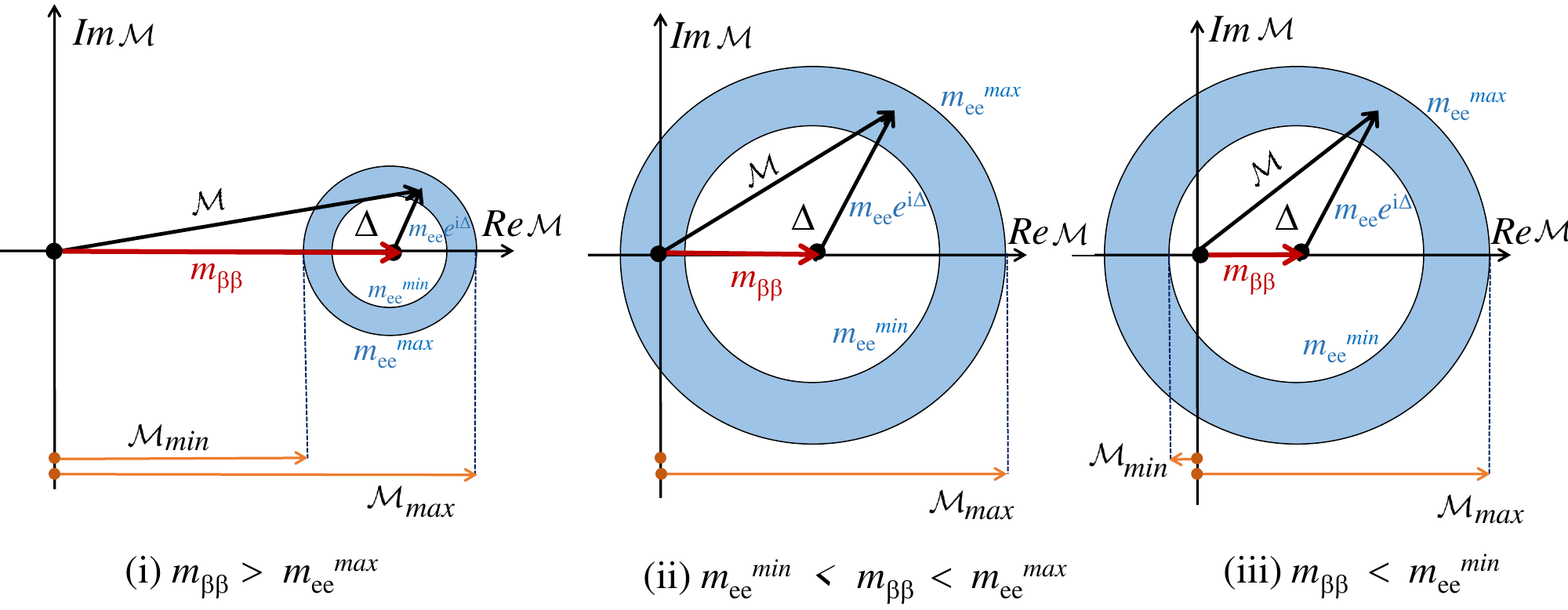}~~
  \caption{The mass ${\cal M}$ in the complex plane for different relative values of $|m_{\beta \beta}|$ and $|m_{ee}|$. The blue ring shows the contribution of light neutrinos with possible uncertainties and varying $\Delta$. The mass ${\cal M}$ is described by a vector which connects the origin with any point inside the ring.}
  \label{schem}
\end{figure*}

Notice that the same result as in Eq.~(\ref{eq:mass4}) can be obtained from the decoupling of the sterile neutrino, which reduces the $4 \times 4$ mass matrix to the $3 \times 3$ matrix by block diagonalization. 
This leads to  the $ee-$element of the $3 \times 3$ matrix 
\begin{equation}
m_{ee} =  m_{ee}^0 - \frac{m_{eS}^2}{m_S} \approx m_{ee}^0 - m_4 e^{i\alpha_4} \tan^2 \theta_{14},   
\label{eq:mass44}
\end{equation}
where $\tan^2 \theta_{14} \approx  |U_{e4}|^2$. 

From Eq.~(\ref{eq:mass4}), we obtain an expression for a combination of sterile neutrino parameters:
\begin{equation}
|U_{e4}|^4 m_4^2 = \left| |m_{ee}^0| -  e^{i\Delta} |m_{ee}| \right|^2. 
\label{eq:ster}
\end{equation}
Here, $\Delta  \equiv \phi - \gamma$ is the relative phase of $m_{ee}^0$ and $m_{ee}$. 

The mass matrix element $m_{ee}^0$ is related to the effective Majorana mass of the electron neutrino 
$m_{\beta\beta}$ that determines the $0\nu \beta\beta$ decay rate. Indeed, the amplitude of the decay is proportional to
\begin{equation}
    A_{\beta \beta} \propto   \frac{1}{q^2} \sum_{i = 1 - 3} (U_{ei}^*)^2 m_i
    + \frac{1}{q^2 - m_4^2} (U_{e4}^*)^2 m_4, 
\label{eq:aplitude}
\end{equation}
where $q \sim 100~{\rm MeV}$ is the 4-momentum of the neutrino. Here, we have neglected the masses of the light neutrinos in the corresponding propagators since $m_i^2 \ll q^2$.  The amplitude can be rewritten as 
\begin{eqnarray}
A_{\beta \beta} & \propto &   \frac{1}{q^2} \sum_{i = 1 - 4} (U_{ei}^*)^2 m_i
    + \left[\frac{1}{q^2 - m_4^2} -\frac{1}{q^2}\right] (U_{e4}^*)^2 m_4 
\nonumber\\
& = &\frac{1}{q^2} m_{ee}^0 + (U_{e4}^*)^2 m_4 \frac{m_4^2}{q^2(q^2 - m_4^2)}. 
\label{eq:aplitude2}
\end{eqnarray}
Defining $m_{\beta\beta}$ in such a way that  $A \propto m_{\beta\beta}/q^2$, we obtain from Eq.~(\ref{eq:aplitude2}) that 
\begin{equation}
\label{main3}
m_{\beta\beta} = m_{ee}^0 + m_4 (U_{e4}^*)^2 \frac{m_4^2}{(q^2 - m_4^2)}.
\end{equation}
Since $q \sim 100$ MeV, the second term  equals  $\sim 10^{-16}{\rm eV} (U_{e4}^*)^2 (m_4 /1 {\rm eV})^3$. 
Being important for $m_4 > 1$ MeV, it is negligible for $m_4 \sim 1$ eV, so that with high accuracy $|m_{ee}^0| = |m_{\beta\beta}|$.

From Eq. (\ref{eq:ster}), we thus obtain
\begin{equation}
\label{relation0}
\Delta m_{41}^2 =  \frac{4 \left| |m_{\beta\beta}| - e^{i\Delta} |m_{ee}| \right|^2
  }{2 - \sin^2 2\theta_{14}  - 2 \sqrt{1 - \sin^2 2\theta_{14}}} , 
\end{equation}
where we used that $|m_4|^2 \approx \Delta m_{41}^2$ and $\sin^2 2\theta_{14} = 4 |U_{e4}|^2 (1 - |U_{e4}|^2)$. For small  $\sin^2 2\theta_{14}$, this reduces to 
\begin{equation}
\label{relation}
\Delta m_{41}^2 \simeq \frac{16}{(\sin^2 2\theta_{14})^2} \left| |m_{\beta\beta}| 
-  e^{i\Delta} |m_{ee}| \right|^2.
\end{equation}
Eqs. (\ref{relation0}) and (\ref{relation}) provide a relation between   \(\Delta m_{41}^2\)  and  the mixing parameter \(\sin^2{2\theta_{14}}\) 
(which has a universal character) with mass parameter 
\begin{equation}
|{\cal M}|  \equiv \left| |m_{\beta\beta}| - e^{i\Delta}|m_{ee}| \right|
\label{eq:mucal}
\end{equation}
that depends on $m_{\beta\beta}$ and the oscillation data encoded in $|m_{ee}|$ and $\Delta$. 

\begin{figure}[tb!]
\centering
  \includegraphics[width=0.48\textwidth]{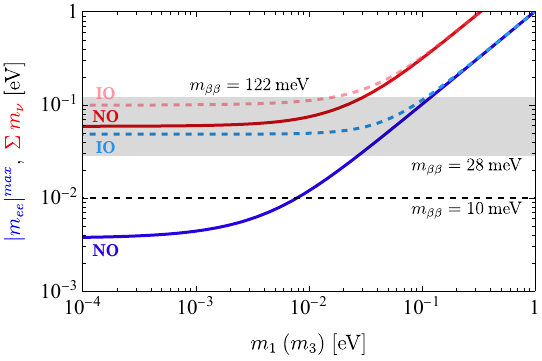}~~
  \caption{Dependence of $|m_{ee}|^{max}$ (blue) and the sum of masses $\sum m_\nu$ (red) on the lightest mass for normal and inverted mass orderings. The bounds on \(m_{\beta\beta}\) are presented for \(m_{\beta\beta} = 122\), \(28\), and \(10\) meV. Solid and dashed lines denote normal and inverted mass hierarchy, respectively.}
  \label{meesum1}
\end{figure}

Notice that an increase of $\Delta m_{41}^2$ with a decrease of the mixing is faster than that along the line of constant oscillation probability for small mass squared difference: $\Delta m_{41}^2 \sim 1/ \sin^2 2\theta_{14}$. 

Let us consider the two quantities $\vert m_{ee}\vert$ and $\Delta$ that enter Eq.~(\ref{relation}) in detail. 
The mass  $|m_{ee}|$ according to Eq.~(\ref{eq:mass3}) is determined by oscillation parameters of active neutrinos. 
For small enough $|U_{e4}|$, one can use for these parameters the results of the fit of the oscillation data in the $3\nu$ framework, ignoring sterile neutrino mixing. A large difference of $\Delta m^2$ allows to disentangle 
the effects of the three light neutrinos from the eV scale neutrino. In fact, $|m_{ee}|$ is determined from experiments 
in which oscillations driven by the fourth neutrino are averaged out, such that their effect can be accounted for, at least partially, by the renormalization of neutrino fluxes. Clearly, this can not affect the determination of $\Delta m_{ij}^2$ and therefore $m_i$. As far as mixing is concerned,  one should consider the impact of sterile neutrinos in experiments that measure the mixing angles $\theta_{12}$ and $\theta_{13}$. In the case of solar neutrinos, the values $|U_{e4}|^2 < 0.05$ \cite{Goldhagen:2021kxe} are allowed. For the KamLAND experiment, the effect of sterile neutrinos can be absorbed in the flux uncertainties. In the case of reactor experiments on 1-3 mixing measurements, results will not be affected if $|U_{e4}|^2$ is smaller than the value extracted from Daya Bay and RENO measurements. For $\Delta m_{41}^2 \gtrsim 1$ eV$^2$, this gives $|U_{e4}|^2 < 0.05$. So,  our results for $|m_{ee}|$ are valid in the range of $|U_{e4}|^2$ below 0.05 and for $|m_{ee}|$, we can use the intervals obtained in the $ 3\nu$ framework \cite{Esteban:2020cvm}.

In addition, the accuracy of the approximation can be estimated by decoupling the sterile neutrino in the mass matrix and considering the corrections to this procedure.

Let us show that by varying the unknown phases $\alpha_i$,  the phase $\Delta$ can take an arbitrary value.  Indeed,  the expression (\ref{eq:mass3}), which defines the phase $\phi$, can be written as 
\begin{equation}
|m_{ee}| e^{i\phi} \equiv e^{i\alpha_1}\sum_{i = 1-3}  |U_{ei}|^2 m_i e^{i(\alpha_i - \alpha_1)}
= e^{i(\alpha_1 + \xi)} |m_{ee}|, 
\label{eq:maeee}
\end{equation}
so  that $\phi = \alpha_1 + \xi$. The phase $\xi$ is determined by the relative phases $(\alpha_2 - \alpha_1)$ and    $(\alpha_3 - \alpha_1)$, and since  $\alpha_2$ and $\alpha_3$ are arbitrary, these relative phases can be arbitrary as well for any given value of $\alpha_1$.  The  phase $\phi = \alpha_1 + \xi$, and therefore $\Delta$ are also arbitrary since $\alpha_1$ is arbitrary. (If $m_1 = 0$, the exponent with the phase $\alpha_2$ should be factorized instead of the exponent with $\alpha_1$). \\

\begin{figure*}[htb!]
\centering
  \includegraphics[width=\textwidth]{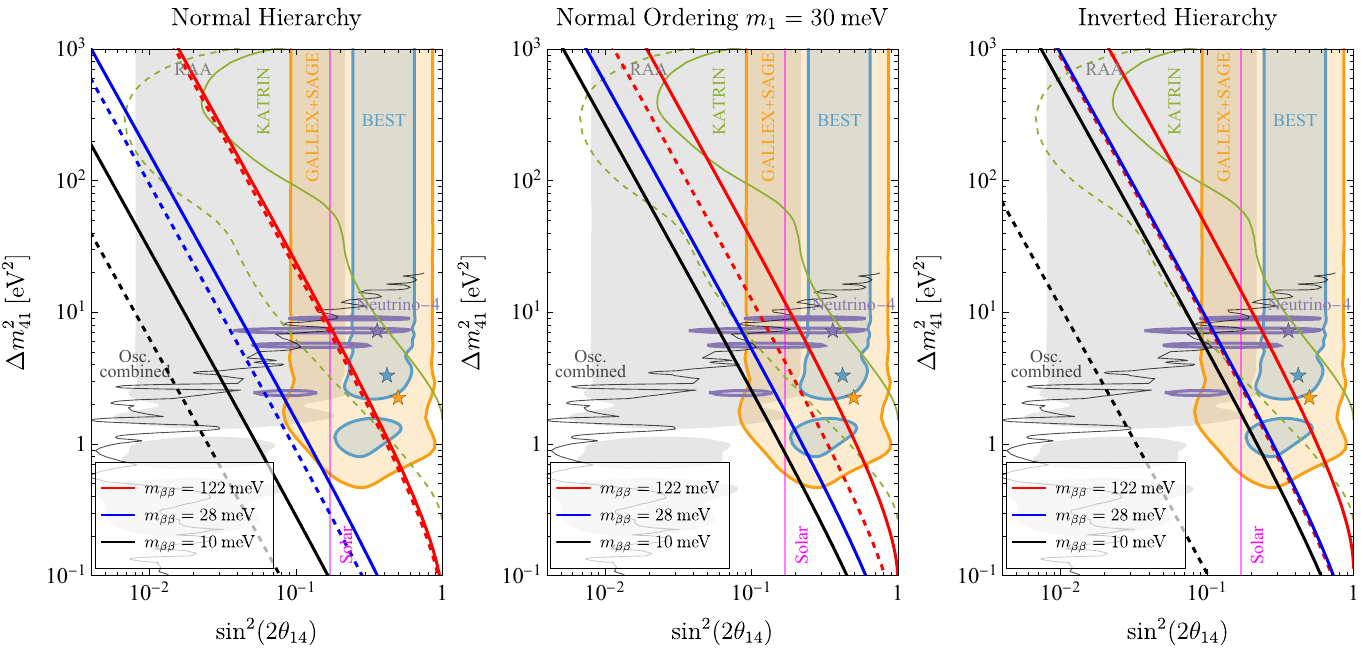}
  \caption{
Upper bounds on  $\sin^22\theta_{14}$ and $\Delta m_{41}^2$ from $0\nu \beta \beta$ decay for different values of $m_{\beta\beta}$, the phase $\Delta$ and type of neutrino mass spectrum. Solid lines show the conservative bound  ($\Delta = \pi$), while the dashed lines indicate the most optimistic bound ($\Delta = 0$).  Red, blue and black lines correspond to $m_{\beta\beta} = 122, ~ 28$ and $10$ meV, respectively. Left panel: normal mass hierarchy, $m_1=0$; central panel: normal mass ordering with $m_1 = 30$ meV; right panel: inverted mass hierarchy, $m_3=0$.  The favored parameter space of the Gallium calibration experiments, BEST, Neutrino-4, and the reactor antineutrino anomaly (RAA) are shown. We also presented the combined bound from oscillation experiments (gray line), the KATRIN (solid green line), and expected KATRIN sensitivity (dashed green lines), as well as the bound from solar neutrino data (purple vertical line). 
}
  \label{money1}
\end{figure*}

\vspace{0.1 in}
\textbf{\emph{Bounds on sterile neutrinos from the  $0\nu\beta\beta$ decay}.-- }
Let us analyze the constraints on parameters of sterile neutrinos that can be obtained from Eqs. (\ref{relation0}) and  (\ref{relation}) and are determined by $|{\cal M}|$.  For $m_{\beta\beta}$, we use the latest constraint  from KamLAND-Zen \cite{Abe:2024rpr}
\begin{equation}
m_{\beta \beta} < (28 - 122)~~ {\rm meV}~~~~ (90\% ~{\rm C.L.}),   
\label{kamland}
\end{equation}
where the interval corresponds to uncertainty in the nuclear matrix element. We also consider the projected upper bound $m_{\beta\beta} \leq 10$ meV from the future LEGEND-1000 experiment \cite{LEGEND:2021bnm}.

In the limit  $|m_{ee}| \ll m_{\beta\beta}$   (realized presently for the normal mass hierarchy) the mass $|m_{ee}|$ can be neglected in 
Eq.~ (\ref{relation}), so that 
\begin{equation}
|{\cal M}|^2 \approx |m_{\beta\beta}|^2.
\label{relation2}
\end{equation}
Therefore the  upper bound on $m_{\beta\beta}$ gives directly  an upper bound on $\Delta m_{41}^2$ as function of $\sin^2 2\theta_{14}$, 
or an  upper bound on $\sin^2 2\theta_{14}$ for a given $\Delta m_{41}^2$. Another extreme  is when $|m_{\beta\beta}| \ll |m_{ee}|$.  Again, the phase $\Delta$ is irrelevant, and we obtain
\begin{equation}
\label{relation3}
|{\cal M}|^2 \approx |m_{ee}|^2.
\end{equation}
This implies strong cancellation between the contributions of the light neutrinos and the eV scale neutrino, that is, equality of the absolute values of the contributions of the three light neutrinos and the eV scale neutrino. In this case, further strengthening of the bound on $m_{\beta\beta}$ will not improve restrictions on parameters of sterile neutrino. Such a situation would be realized in the case of inverted mass hierarchy, which predicts $|m_{ee}| = (30 - 100)$ meV. 

For 
\begin{equation}
|m_{\beta\beta}| = |m_{ee}|, ~~~~ \Delta = 0,
\label{eq:cancell}
\end{equation}
the cancellation occurs and $|{\cal M}|^2 = 0$. The cancellation implies  $\sin^2 2\theta_{14} \rightarrow 0$,  that is,  no room for sterile neutrino contribution. 

In general, the bound depends on $|m_{ee}|$ and the arbitrary phase $\Delta$. Varying $\Delta$, we get an interval of bounds on $\Delta m^2_{41}$, which is determined by relative values of $m_{\beta\beta}$, $|m_{ee}|^{max}$ and $|m_{ee}|^{min}$ -- maximal and minimal values of $|m_{ee}|$. The edges of this interval are given by $\Delta = \pi$ and $0$. The conservative limit on $\Delta m_{41}^2$, which is valid for all the cases considered below, corresponds to the plus sign in $|{\cal M}|$ ($\Delta = \pi$): 
\begin{equation}
\label{relation1x}
\Delta m_{41}^2 = \frac{16}{(\sin^2 2\theta_{14})^2} \left| |m_{\beta\beta}| + |m_{ee}|^{max} \right|^2.
\end{equation}

There are three possibilities (see Fig.~\ref{schem}):

\begin{itemize}
\item
$|m_{\beta\beta}| > |m_{ee}|^{max}$ (diagram (i)): the interval is given by $|m_{\beta\beta}| \pm  |m_{ee}|^{max}$. 
The conservative limit is  (\ref{relation1x})
($\Delta = \pi$):

\item
$|m_{ee}|^{max}  >   |m_{\beta\beta}| > |m_{ee}|^{min}$ (diagram (ii)): 
 the conservative bound is the same as in Eq.~\ref{relation1x}. 
Now under condition (\ref{eq:cancell}) complete cancellation between $|m_{\beta\beta}|$ and  $|m_{ee}|$ is possible 
in Eq.~(\ref{relation})   
which leads to the absence of the lower edge of the bound.

\item 
$|m_{\beta\beta}| < |m_{ee}|^{min}$ (diagram (iii)):
the cancellation  (\ref{eq:cancell})  is not realized and the lower edge
of the interval is given by 
$$
|{\cal M}| =  |m_{ee}|^{min} - |m_{\beta\beta}|.
$$
\end{itemize}

In Fig. \ref{meesum1}, we show the dependence of $|m_{ee}|^{max}$  as well as the sum of masses, $\sum m_\nu = m_1 + m_2 + m_3$, on the value of the lightest mass for normal and inverted mass orderings. The maximal value of $|m_{ee}|$ corresponds to zero relative phases of light neutrinos, that is $\alpha_1 = \alpha_2 = \alpha_3$. The solid and dashed blue lines represent the $|m_{ee}|^{max}$ allowed by oscillation experiments for normal and inverted ordering, respectively. The solid and dashed red lines represent $\sum m_\nu$ for normal and inverted ordering, respectively. This figure allows one to find $|m_{ee}|^{max}$ and check that there is no contradiction to the cosmological bound. 
The current bound $\sum m_\nu \leq$ 0.12 eV \cite{Planck:2018vyg}, which coincides with the upper bound from KAMLAND-Zen, allows the lightest neutrino mass in normal ordering to be up to 30 meV. The gray shaded band and black dashed line indicate the current and projected future bounds from \( 0\nu \beta \beta \).  

Let us apply this consideration to the present bounds on $m_{\beta\beta}$ and light neutrino parameters.
In Fig~\ref{money1}, we show the bounds on sterile neutrinos in the $\Delta m^2_{41} - \sin^2 2\theta_{14}$ plane for 
the benchmark values of $m_{\beta\beta} = $ 122, 28, and 10 meV and different mass spectra of light neutrinos. 

\begin{figure*}[htb!]
\centering
\includegraphics[width=\textwidth]{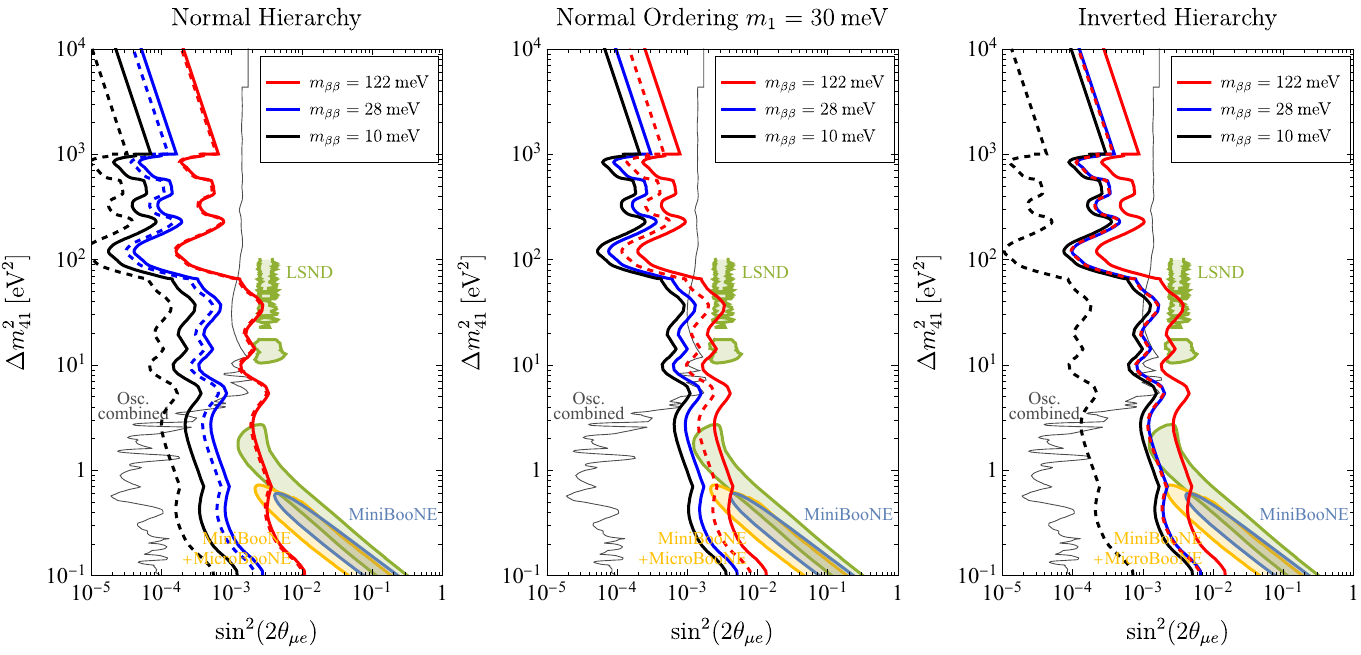}
  \caption{
Upper bounds on parameters of sterile neutrino in the $\sin^22\theta_{\mu e}-\Delta m_{41}^2$ plane 
from the $0\nu \beta \beta$  decay and $\nu_\mu$ disappearance experiments. 
Solid lines show the conservative bound  ($\Delta = \pi$), while the dashed lines indicate the most optimistic bound ($\Delta = 0$).  
Red, blue and black lines  correspond to $m_{\beta\beta} = 122, ~ 28$ and $10$ meV, respectively. 
Left panel: normal mass hierarchy, $m_1=0$; central panel: normal mass ordering 
with $m_1 = 30$ meV; right panel: inverted mass hierarchy, $m_3=0$. 
Shown are favored regions of parameters from the LSND, MiniBooNE, and combined MiniBooNE and MicroBooNE experiments, as well as the bounds from oscillation experiments.  
}
\label{money2}
\end{figure*}

For normal mass hierarchy, with $m_1 = 0$ (Fig.~\ref{money1} left panel), the mass $|m_{ee}|^{max} = 3.7$ meV (corresponding to zero relative phases between contributions) is much smaller than the bound on $m_{\beta \beta}$. So, configuration (i) is realized for all three benchmark values of $m_{\beta\beta}$. With decrease of $m_{\beta \beta}$ the width of the interval characterized by the ratio $r$ of $\Delta m_{41}^2$ in the upper and low edges increases:  $r = 1.13,~ 1.70, ~4.73$ for 122, 28 and 10 meV correspondingly. One can extrapolate the conservative bound to larger $\Delta m^2_{41}$ using the following numerical expression: $\Delta m^2_{41} = 25.3\, \textrm{eV}^2/(\sin^2 2\theta_{14})^2.$

In the case of a non-hierarchical spectrum with normal ordering and $m_1 = 30$ meV (middle panel of Fig.~\ref{money1}), we obtain $m_2 = 31.2$ meV and $m_3 = 58.4$ meV for the two other masses. The sum of masses $\sim 120$ meV is at the upper cosmological bound \cite{Planck:2018vyg}. Then, according to Fig. 1, $|m_{ee}|^{max} = 31.0$ meV. The minimal value is $|m_{ee}|^{min} \approx 9.6$ meV. For $m_{\beta \beta} = 122$ meV, the configuration (i) is realized so that both higher and lower edges exist with $r =  2.83$. For $|m_{\beta \beta}| = 28$ and 10 meV, we have a configuration (ii). Now, cancellation is possible, and only the higher, conservative edge exists.

In the case of the inverted hierarchy with $m_3 = 0$ (Fig.~\ref{money1} right panel), we have $|m_{ee}|^{max} = 48.3$ meV and $|m_{ee}|^{min} = 18.3$ meV. For  $|m_{\beta \beta}| = 122$ meV, the configuration (i) is realized
and we find $r =$ 5.34. For $|m_{\beta \beta}| = 28$ meV, the case (ii) is realized with only the conservative bound and the possibility of cancellation (cf. Eq.~\ref{eq:cancell}). The future bound $|m_{\beta \beta}| = 10$ meV would correspond to configuration (iii). Now, both higher and lower edges are present.

Comparing results shown in panels of Fig.~\ref{money1}, we see that the conservative bound slightly weakens from normal hierarchy (NH) and normal ordering (NO) to inverted hierarchy (IH) for  $m_{\beta\beta} = 122$ meV which dominates in ${\cal M}$. With the improvement of the bound, this weakening becomes more substantial. E.g., for  $m_{\beta\beta} = 28$ meV and   $\Delta m^2_{41} = 10$ eV$^2$, the upper bounds on $\sin^2 2 \theta_{14}$ worsen as  $0.04, ~ 0.07, ~ 0.10$ for NH, NO and IH correspondingly.

Notice that the phase $\Delta$ could be determined if both $0\nu\beta \beta$ and sterile neutrinos are discovered.  Or if the theory of neutrino mass and mixing is constructed, which will allow to predict the value of $\Delta$. For example, if a theory predicts zero values of the Majorana phases and the Dirac phase equals $\pi$  we have  $\Delta = 0$. Then, in the case of non-hierarchical spectrum with normal ordering and $m_1 = 30$ meV, the value $|m_{ee}|=$ 28.4 meV is obtained. With this, $m_{\beta\beta} = 122$ meV leads to the two times stronger bound on $\sin^2 2 \theta_{14}$ that for unfixed phases (red line in Fig.~\ref{money1}, middle). Until that, only the conservative bound makes sense, and variations of the bound with $\Delta$  should be treated as possible uncertainty.

In Fig. ~\ref{money1}, we also show the regions of sterile neutrino parameters suggested by the Gallium 
anomaly \cite{Giunti:2010zu}, reactor antineutrino anomaly \cite{Mention:2011rk} along with the results from BEST \cite{Barinov:2022wfh} and Neutrino-4 \cite{Serebrov:2020kmd} experiments. Bounds on sterile neutrino parameters are imposed by the reactor experiments DANSS  \cite{Danilov:2022bss}, Daya Bay \cite{DayaBay:2024nip}, PROSPECT~\cite{Andriamirado:2024glk}, RENO and NEOS \cite{RENO:2020hva}, and STEREO~ \cite{STEREO:2022nzk}. 
The most updated and conservative combined bound from these experiments is shown by the gray solid line. 
We also show the bounds from the solar neutrino data assuming the GS98 solar model \cite{Goldhagen:2021kxe} and from \(^3\)H decay experiments KATRIN \cite{KATRIN:2022ith},  for which we also include the projected final sensitivity (1000 days at nominal column density) \cite{KATRIN:2020dpx, KATRIN:2022ith}. Confronting our bounds from $0\nu \beta\beta$ decay with these positive and negative results, we find the following.

According to Fig. ~\ref{money1},  the \(0\nu\beta\beta\) data impose much tighter constraints than the oscillation and solar neutrino data for $\Delta m_{41}^2\geq 10 $ eV$^2$. The sterile neutrino parameter space implied by the Gallium anomaly, while not entirely excluded by oscillation experiments alone, is completely ruled out by combined data from oscillation and \(0\nu\beta\beta\) experiments. Almost the entire BEST region (including the best-fit point) is ruled out by the conservative $0\nu \beta\beta$ bound. The high energy part of the Neutrino-4 region with the best-fit point is ruled out by the $m_{\beta\beta} = 122$ meV bound and completely ruled out by $m_{\beta\beta} = 28$ meV. With the latter value the $0\nu \beta\beta$ decay gives a better bound than other experiments for $\Delta m^2_{41} > (4 - 6)$ eV$^2$. Notably, Fig. ~\ref{money1} shows that  \(0\nu\beta\beta\) bound is much stronger than the present KATRIN bound for $\Delta m_{41}^2\geq 10 $ eV$^2$. For normal hierarchy, KATRIN's projected sensitivity \cite{KATRIN:2020dpx, KATRIN:2022ith} is weaker across the entire mass range compared to the  $m_{\beta\beta} = 28$ meV bound of KamLAND-Zen.

\vspace{0.1 in}
{\textbf {\textit {Bounds on the $\nu_\mu - \nu_e$ channel from the ${0\nu\beta\beta}$.--}}}
The bound on $|U_{e4}|$ as a function of $m_4$ obtained from  ${0\nu\beta\beta}$ can be used together 
with the bound on  $|U_{\mu4}|$ from disappearance accelerator and atmospheric neutrino experiments to restrict short 
baseline $\nu_\mu - \nu_e$ oscillations claimed by LSND \cite{LSND:2001aii}, MiniBooNE \cite{MiniBooNE:2018esg, MiniBooNE:2020pnu} and combined MiniBooNE and MicroBooNE experiments \cite{MiniBooNE:2022emn}. 
Recall that the effective mixing parameter for the $\nu_\mu \to \nu_e$ appearance is given by
\begin{align}
    \sin^22\theta_{\mu e}  = 4\vert U_{e 4}(\Delta m_{41}^2) \vert^2 \vert U_{\mu 4} (\Delta m_{41}^2)\vert^2 . 
\label{eq:emu}
\end{align}
The limits on $\vert U_{\mu 4}\vert^2$ are taken from the CCFR  \cite{Stockdale:1984cg}, CDHS \cite{Dydak:1983zq}, IceCube  \cite{IceCube:2016rnb}, IceCube DeepCore \cite{Abbasi:2024ktc}, MINOS(+) 
\cite{MINOS:2017cae}, Super-Kamiokande \cite{Super-Kamiokande:2014ndf}, KARMEN \cite{KARMEN:2002zcm} and NOMAD \cite{NOMAD:2003mqg} experiments. 
We use the relation (\ref{eq:emu}) to obtain the bound on $\sin^22\theta_{\mu e} $ as a function of $\Delta m_{41}^2$.

In Fig.~\ref{money2} we present the bounds from ${0\nu\beta\beta}$ decay in the $\sin^22\theta_{\mu e}-\Delta m_{41}^2$ plane. 
Also shown are the regions of parameters that can explain the LSND and MiniBooNE anomalies.  For comparison, we present 
the bounds (the gray solid line) obtained by combining the restrictions on $\vert U_{e4}\vert^2$ and $\vert U_{\mu 4}\vert^2$ from oscillation experiments. 
The bounds have similar characteristics as in Fig.~\ref{money1} with the $0\nu \beta\beta$ decay bound dominating  at $\Delta m^2_{41} > (4 - 10)$ eV$^2.$
According to Fig.~\ref{money2}, the parameter space favored by MiniBooNE data is entirely excluded by current \(0\nu\beta\beta\) decay data. 
The parameter space favored by LSND is almost entirely ruled out by the \(m_{\beta\beta}\) bound of 122 meV. 
The optimistic bound $m_{\beta\beta} \leq$ 28 meV excludes the entire parameter space favored by LSND, MiniBooNE, 
and the combined MiniBooNE and MicroBooNE experiments for both normal and inverted mass hierarchies. 
With a future \(m_{\beta\beta}\) bound of 10 meV, the restriction, especially for normal hierarchy, becomes even more stringent.

\vspace{0.1 in}
{\textbf {\textit {Conclusions.--}}} 
In view of the latest KamLAND-Zen results on $0\nu \beta\beta$  decay and new cosmological bound on the sum of neutrino masses, we refined the restrictions on sterile neutrinos. We presented a compact formula for the $0\nu \beta\beta$ decay bounds in the  $\Delta m^2_{41} - \sin^2 2\theta_{14}$ plane which depends on relevant parameters only. The formula is convenient to analyze the present and future experimental results. We explored in detail the dependency of the bounds on unknown CP-phases and the mass spectra of light neutrinos, including their mass ordering and potential degeneracy. The bounds we have derived are more stringent than those found in prior studies.

We have found that the bound from $0\nu \beta\beta$ decay dominates over the oscillation bounds at $\Delta m^2_{41} > 6 - 10$ eV$^2$. We show that sterile neutrinos at the eV mass scale, prompted by the Gallium, BEST, and Neutrino-4 anomalies, are under tension with the present conservative bound of 122 meV from KamLAND-Zen, and they are ruled by the optimistic bound of $28$ meV. For the first time, we demonstrate that the sterile neutrino parameter space consistent with the LSND and MiniBooNE anomalies are tightly constrained and, in most cases, ruled out by $0\nu \beta\beta$  decay result in combination with $\nu_\mu-$ disappearance.

\vspace{0.1 in}
{\textbf {\textit {Acknowledgments.--}}} SJ thanks K.S. Babu and Raj Gandhi for useful discussions. SJ also wishes to acknowledge the Center for Theoretical Underground Physics and Related Areas (CETUP*) and the Institute for Underground Science at SURF for their warm hospitality and for providing a stimulating environment. 

\bibliographystyle{utcaps_mod}
\bibliography{reference}

\providecommand{\href}[2]{#2}\begingroup\raggedright\begin{thebibliography}{10}

\bibitem{LSND:2001aii}
{\normalfont \bfseries LSND}, A.~Aguilar {\em et al.}, ``{\em {Evidence for neutrino oscillations from the observation of $\bar{\nu}_e$ appearance in a $\bar{\nu}_\mu$ beam}},'' \href{http://dx.doi.org/10.1103/PhysRevD.64.112007}{Phys. Rev. D {\normalfont \bfseries 64} (2001)  112007}, \href{http://arxiv.org/abs/hep-ex/0104049}{{\normalfont \ttfamily arXiv:hep-ex/0104049}}.

\bibitem{MiniBooNE:2020pnu}
{\normalfont \bfseries MiniBooNE}, A.~A. Aguilar-Arevalo {\em et al.}, ``{\em {Updated MiniBooNE neutrino oscillation results with increased data and new background studies}},'' \href{http://dx.doi.org/10.1103/PhysRevD.103.052002}{Phys. Rev. D {\normalfont \bfseries 103} (2021) no.~5, 052002}, \href{http://arxiv.org/abs/2006.16883}{{\normalfont \ttfamily arXiv:2006.16883}}.

\bibitem{GALLEX:1997lja}
{\normalfont \bfseries GALLEX}, W.~Hampel {\em et al.}, ``{\em {Final results of the Cr-51 neutrino source experiments in GALLEX}},'' \href{http://dx.doi.org/10.1016/S0370-2693(97)01562-1}{Phys. Lett. B {\normalfont \bfseries 420} (1998)  114--126}.

\bibitem{Kaether:2010ag}
F.~Kaether, W.~Hampel, G.~Heusser, J.~Kiko, and T.~Kirsten, ``{\em {Reanalysis of the GALLEX solar neutrino flux and source experiments}},'' \href{http://dx.doi.org/10.1016/j.physletb.2010.01.030}{Phys. Lett. B {\normalfont \bfseries 685} (2010)  47--54}, \href{http://arxiv.org/abs/1001.2731}{{\normalfont \ttfamily arXiv:1001.2731}}.

\bibitem{SAGE:1998fvr}
{\normalfont \bfseries SAGE}, J.~N. Abdurashitov {\em et al.}, ``{\em {Measurement of the response of the Russian-American gallium experiment to neutrinos from a Cr-51 source}},'' \href{http://dx.doi.org/10.1103/PhysRevC.59.2246}{Phys. Rev. C {\normalfont \bfseries 59} (1999)  2246--2263}, \href{http://arxiv.org/abs/hep-ph/9803418}{{\normalfont \ttfamily arXiv:hep-ph/9803418}}.

\bibitem{Abdurashitov:2005tb}
J.~N. Abdurashitov {\em et al.}, ``{\em {Measurement of the response of a Ga solar neutrino experiment to neutrinos from an Ar-37 source}},'' \href{http://dx.doi.org/10.1103/PhysRevC.73.045805}{Phys. Rev. C {\normalfont \bfseries 73} (2006)  045805}, \href{http://arxiv.org/abs/nucl-ex/0512041}{{\normalfont \ttfamily arXiv:nucl-ex/0512041}}.

\bibitem{Acero:2007su}
M.~A. Acero, C.~Giunti, and M.~Laveder, ``{\em {Limits on nu(e) and anti-nu(e) disappearance from Gallium and reactor experiments}},'' \href{http://dx.doi.org/10.1103/PhysRevD.78.073009}{Phys. Rev. D {\normalfont \bfseries 78} (2008)  073009}, \href{http://arxiv.org/abs/0711.4222}{{\normalfont \ttfamily arXiv:0711.4222}}.

\bibitem{Giunti:2010zu}
C.~Giunti and M.~Laveder, ``{\em {Statistical Significance of the Gallium Anomaly}},'' \href{http://dx.doi.org/10.1103/PhysRevC.83.065504}{Phys. Rev. C {\normalfont \bfseries 83} (2011)  065504}, \href{http://arxiv.org/abs/1006.3244}{{\normalfont \ttfamily arXiv:1006.3244}}.

\bibitem{Kostensalo:2019vmv}
J.~Kostensalo, J.~Suhonen, C.~Giunti, and P.~C. Srivastava, ``{\em {The gallium anomaly revisited}},'' \href{http://dx.doi.org/10.1016/j.physletb.2019.06.057}{Phys. Lett. B {\normalfont \bfseries 795} (2019)  542--547}, \href{http://arxiv.org/abs/1906.10980}{{\normalfont \ttfamily arXiv:1906.10980}}.

\bibitem{Mention:2011rk}
G.~Mention, M.~Fechner, T.~Lasserre, T.~A. Mueller, D.~Lhuillier, M.~Cribier, and A.~Letourneau, ``{\em {The Reactor Antineutrino Anomaly}},'' \href{http://dx.doi.org/10.1103/PhysRevD.83.073006}{Phys. Rev. D {\normalfont \bfseries 83} (2011)  073006}, \href{http://arxiv.org/abs/1101.2755}{{\normalfont \ttfamily arXiv:1101.2755}}.

\bibitem{NEUTRINO-4:2018huq}
{\normalfont \bfseries NEUTRINO-4}, A.~P. Serebrov {\em et al.}, ``{\em {First Observation of the Oscillation Effect in the Neutrino-4 Experiment on the Search for the Sterile Neutrino}},'' \href{http://dx.doi.org/10.1134/S0021364019040040}{Pisma Zh. Eksp. Teor. Fiz. {\normalfont \bfseries 109} (2019) no.~4, 209--218}, \href{http://arxiv.org/abs/1809.10561}{{\normalfont \ttfamily arXiv:1809.10561}}.

\bibitem{Serebrov:2020kmd}
A.~P. Serebrov {\em et al.}, ``{\em {Search for sterile neutrinos with the Neutrino-4 experiment and measurement results}},'' \href{http://dx.doi.org/10.1103/PhysRevD.104.032003}{Phys. Rev. D {\normalfont \bfseries 104} (2021) no.~3, 032003}, \href{http://arxiv.org/abs/2005.05301}{{\normalfont \ttfamily arXiv:2005.05301}}.

\bibitem{Barinov:2022wfh}
V.~V. Barinov {\em et al.}, ``{\em {Search for electron-neutrino transitions to sterile states in the BEST experiment}},'' \href{http://dx.doi.org/10.1103/PhysRevC.105.065502}{Phys. Rev. C {\normalfont \bfseries 105} (2022) no.~6, 065502}, \href{http://arxiv.org/abs/2201.07364}{{\normalfont \ttfamily arXiv:2201.07364}}.

\bibitem{Danilov:2022bss}
{\normalfont \bfseries DANSS}, M.~Danilov, ``{\em {New results from the DANSS experiment}},'' \href{http://dx.doi.org/10.22323/1.414.0616}{PoS {\normalfont \bfseries ICHEP2022} (2022)  616}, \href{http://arxiv.org/abs/2211.01208}{{\normalfont \ttfamily arXiv:2211.01208}}.

\bibitem{DayaBay:2024nip}
{\normalfont \bfseries Daya Bay}, F.~P. An {\em et al.}, ``{\em {Search for a sub-eV sterile neutrino using Daya Bay's full dataset}},'' \href{http://arxiv.org/abs/2404.01687}{{\normalfont \ttfamily arXiv:2404.01687}}.

\bibitem{Andriamirado:2024glk}
M.~Andriamirado {\em et al.}, ``{\em {Final Search for Short-Baseline Neutrino Oscillations with the PROSPECT-I Detector at HFIR}},'' \href{http://arxiv.org/abs/2406.10408}{{\normalfont \ttfamily arXiv:2406.10408}}.

\bibitem{RENO:2020hva}
{\normalfont \bfseries RENO, NEOS}, Z.~Atif {\em et al.}, ``{\em {Search for sterile neutrino oscillations using RENO and NEOS data}},'' \href{http://dx.doi.org/10.1103/PhysRevD.105.L111101}{Phys. Rev. D {\normalfont \bfseries 105} (2022) no.~11, L111101}, \href{http://arxiv.org/abs/2011.00896}{{\normalfont \ttfamily arXiv:2011.00896}}.

\bibitem{STEREO:2022nzk}
{\normalfont \bfseries STEREO}, H.~Almaz\'an {\em et al.}, ``{\em {STEREO neutrino spectrum of $^{235}$U fission rejects sterile neutrino hypothesis}},'' \href{http://dx.doi.org/10.1038/s41586-022-05568-2}{Nature {\normalfont \bfseries 613} (2023) no.~7943, 257--261}, \href{http://arxiv.org/abs/2210.07664}{{\normalfont \ttfamily arXiv:2210.07664}}.

\bibitem{KATRIN:2022ith}
{\normalfont \bfseries KATRIN}, M.~Aker {\em et al.}, ``{\em {Improved eV-scale sterile-neutrino constraints from the second KATRIN measurement campaign}},'' \href{http://dx.doi.org/10.1103/PhysRevD.105.072004}{Phys. Rev. D {\normalfont \bfseries 105} (2022) no.~7, 072004}, \href{http://arxiv.org/abs/2201.11593}{{\normalfont \ttfamily arXiv:2201.11593}}.

\bibitem{Abazajian:2012ys}
K.~N. Abazajian {\em et al.}, ``{\em {Light Sterile Neutrinos: A White Paper}},'' \href{http://arxiv.org/abs/1204.5379}{{\normalfont \ttfamily arXiv:1204.5379}}.

\bibitem{Acero:2022wqg}
M.~A. Acero {\em et al.}, ``{\em {White Paper on Light Sterile Neutrino Searches and Related Phenomenology}},'' \href{http://arxiv.org/abs/2203.07323}{{\normalfont \ttfamily arXiv:2203.07323}}.

\bibitem{Balaji:2001ns}
K.~R.~S. Balaji, A.~Perez-Lorenzana, and A.~Y. Smirnov, ``{\em {Large lepton mixings induced by sterile neutrino}},'' \href{http://dx.doi.org/10.1016/S0370-2693(01)00547-0}{Phys. Lett. B {\normalfont \bfseries 509} (2001)  111--119}, \href{http://arxiv.org/abs/hep-ph/0101005}{{\normalfont \ttfamily arXiv:hep-ph/0101005}}.

\bibitem{our}
S.~Jana, L.~Puetter, and A.~Y. Smirnov to appear.

\bibitem{Benes:2005hn}
P.~Benes, A.~Faessler, F.~Simkovic, and S.~Kovalenko, ``{\em {Sterile neutrinos in neutrinoless double beta decay}},'' \href{http://dx.doi.org/10.1103/PhysRevD.71.077901}{Phys. Rev. D {\normalfont \bfseries 71} (2005)  077901}, \href{http://arxiv.org/abs/hep-ph/0501295}{{\normalfont \ttfamily arXiv:hep-ph/0501295}}.

\bibitem{Goswami:2005ng}
S.~Goswami and W.~Rodejohann, ``{\em {Constraining mass spectra with sterile neutrinos from neutrinoless double beta decay, tritium beta decay and cosmology}},'' \href{http://dx.doi.org/10.1103/PhysRevD.73.113003}{Phys. Rev. D {\normalfont \bfseries 73} (2006)  113003}, \href{http://arxiv.org/abs/hep-ph/0512234}{{\normalfont \ttfamily arXiv:hep-ph/0512234}}.

\bibitem{Li:2011ss}
Y.~F. Li and S.-s. Liu, ``{\em {Vanishing effective mass of the neutrinoless double beta decay including light sterile neutrinos}},'' \href{http://dx.doi.org/10.1016/j.physletb.2011.11.054}{Phys. Lett. B {\normalfont \bfseries 706} (2012)  406--411}, \href{http://arxiv.org/abs/1110.5795}{{\normalfont \ttfamily arXiv:1110.5795}}.

\bibitem{Barry:2011wb}
J.~Barry, W.~Rodejohann, and H.~Zhang, ``{\em {Light Sterile Neutrinos: Models and Phenomenology}},'' \href{http://dx.doi.org/10.1007/JHEP07(2011)091}{JHEP {\normalfont \bfseries 07} (2011)  091}, \href{http://arxiv.org/abs/1105.3911}{{\normalfont \ttfamily arXiv:1105.3911}}.

\bibitem{Giunti:2012tn}
C.~Giunti, M.~Laveder, Y.~F. Li, Q.~Y. Liu, and H.~W. Long, ``{\em {Update of Short-Baseline Electron Neutrino and Antineutrino Disappearance}},'' \href{http://dx.doi.org/10.1103/PhysRevD.86.113014}{Phys. Rev. D {\normalfont \bfseries 86} (2012)  113014}, \href{http://arxiv.org/abs/1210.5715}{{\normalfont \ttfamily arXiv:1210.5715}}.

\bibitem{Girardi:2013zra}
I.~Girardi, A.~Meroni, and S.~T. Petcov, ``{\em {Neutrinoless Double Beta Decay in the Presence of Light Sterile Neutrinos}},'' \href{http://dx.doi.org/10.1007/JHEP11(2013)146}{JHEP {\normalfont \bfseries 11} (2013)  146}, \href{http://arxiv.org/abs/1308.5802}{{\normalfont \ttfamily arXiv:1308.5802}}.

\bibitem{Pascoli:2013fiz}
S.~Pascoli, M.~Mitra, and S.~Wong, ``{\em {Effect of cancellation in neutrinoless double beta decay}},'' \href{http://dx.doi.org/10.1103/PhysRevD.90.093005}{Phys. Rev. D {\normalfont \bfseries 90} (2014) no.~9, 093005}, \href{http://arxiv.org/abs/1310.6218}{{\normalfont \ttfamily arXiv:1310.6218}}.

\bibitem{Guzowski:2015saa}
P.~Guzowski, L.~Barnes, J.~Evans, G.~Karagiorgi, N.~McCabe, and S.~Soldner-Rembold, ``{\em {Combined limit on the neutrino mass from neutrinoless double-\ensuremath{\beta} decay and constraints on sterile Majorana neutrinos}},'' \href{http://dx.doi.org/10.1103/PhysRevD.92.012002}{Phys. Rev. D {\normalfont \bfseries 92} (2015) no.~1, 012002}, \href{http://arxiv.org/abs/1504.03600}{{\normalfont \ttfamily arXiv:1504.03600}}.

\bibitem{Giunti:2015kza}
C.~Giunti and E.~M. Zavanin, ``{\em {Predictions for Neutrinoless Double-Beta Decay in the 3+1 Sterile Neutrino Scenario}},'' \href{http://dx.doi.org/10.1007/JHEP07(2015)171}{JHEP {\normalfont \bfseries 07} (2015)  171}, \href{http://arxiv.org/abs/1505.00978}{{\normalfont \ttfamily arXiv:1505.00978}}.

\bibitem{Jang:2018zug}
C.~H. Jang, B.~J. Kim, Y.~J. Ko, and K.~Siyeon, ``{\em {Neutrinoless Double Beta Decay and Light Sterile Neutrino}},'' \href{http://dx.doi.org/10.3938/jkps.73.1625}{J. Korean Phys. Soc. {\normalfont \bfseries 73} (2018) no.~11, 1625--1630}, \href{http://arxiv.org/abs/1811.09957}{{\normalfont \ttfamily arXiv:1811.09957}}.

\bibitem{Huang:2019qvq}
G.-Y. Huang and S.~Zhou, ``{\em {Impact of an eV-mass sterile neutrino on the neutrinoless double-beta decays: A Bayesian analysis}},'' \href{http://dx.doi.org/10.1016/j.nuclphysb.2019.114691}{Nucl. Phys. B {\normalfont \bfseries 945} (2019)  114691}, \href{http://arxiv.org/abs/1902.03839}{{\normalfont \ttfamily arXiv:1902.03839}}.

\bibitem{Majhi:2019hdj}
R.~Majhi, C.~Soumya, and R.~Mohanta, ``{\em {Light sterile neutrinos and their implications on currently running long-baseline and neutrinoless double beta decay experiments}},'' \href{http://dx.doi.org/10.1088/1361-6471/ab9797}{J. Phys. G {\normalfont \bfseries 47} (2020) no.~9, 095002}, \href{http://arxiv.org/abs/1911.10952}{{\normalfont \ttfamily arXiv:1911.10952}}.

\bibitem{Goldhagen:2021kxe}
K.~Goldhagen, M.~Maltoni, S.~E. Reichard, and T.~Schwetz, ``{\em {Testing sterile neutrino mixing with present and future solar neutrino data}},'' \href{http://dx.doi.org/10.1140/epjc/s10052-022-10052-2}{Eur. Phys. J. C {\normalfont \bfseries 82} (2022) no.~2, 116}, \href{http://arxiv.org/abs/2109.14898}{{\normalfont \ttfamily arXiv:2109.14898}}.

\bibitem{Esteban:2020cvm}
I.~Esteban, M.~C. Gonzalez-Garcia, M.~Maltoni, T.~Schwetz, and A.~Zhou, ``{\em {The fate of hints: updated global analysis of three-flavor neutrino oscillations}},'' \href{http://dx.doi.org/10.1007/JHEP09(2020)178}{JHEP {\normalfont \bfseries 09} (2020)  178}, \href{http://arxiv.org/abs/2007.14792}{{\normalfont \ttfamily arXiv:2007.14792}}.

\bibitem{Abe:2024rpr}
S.~Abe {\em et al.}, ``{\em {Search for Majorana Neutrinos with the Complete KamLAND-Zen Dataset}},'' \href{http://arxiv.org/abs/2406.11438}{{\normalfont \ttfamily arXiv:2406.11438}}.

\bibitem{LEGEND:2021bnm}
{\normalfont \bfseries LEGEND}, N.~Abgrall {\em et al.}, ``{\em {The Large Enriched Germanium Experiment for Neutrinoless $\beta\beta$ Decay}: {LEGEND-1000 Preconceptual Design Report}},'' \href{http://arxiv.org/abs/2107.11462}{{\normalfont \ttfamily arXiv:2107.11462}}.

\bibitem{Planck:2018vyg}
{\normalfont \bfseries Planck}, N.~Aghanim {\em et al.}, ``{\em {Planck 2018 results. VI. Cosmological parameters}},'' \href{http://dx.doi.org/10.1051/0004-6361/201833910}{Astron. Astrophys. {\normalfont \bfseries 641} (2020)  A6}, \href{http://arxiv.org/abs/1807.06209}{{\normalfont \ttfamily arXiv:1807.06209}}. [Erratum: Astron.Astrophys. 652, C4 (2021)].

\bibitem{KATRIN:2020dpx}
{\normalfont \bfseries KATRIN}, M.~Aker {\em et al.}, ``{\em {Bound on 3+1 Active-Sterile Neutrino Mixing from the First Four-Week Science Run of KATRIN}},'' \href{http://dx.doi.org/10.1103/PhysRevLett.126.091803}{Phys. Rev. Lett. {\normalfont \bfseries 126} (2021) no.~9, 091803}, \href{http://arxiv.org/abs/2011.05087}{{\normalfont \ttfamily arXiv:2011.05087}}.

\bibitem{MiniBooNE:2018esg}
{\normalfont \bfseries MiniBooNE}, A.~A. Aguilar-Arevalo {\em et al.}, ``{\em {Significant Excess of ElectronLike Events in the MiniBooNE Short-Baseline Neutrino Experiment}},'' \href{http://dx.doi.org/10.1103/PhysRevLett.121.221801}{Phys. Rev. Lett. {\normalfont \bfseries 121} (2018) no.~22, 221801}, \href{http://arxiv.org/abs/1805.12028}{{\normalfont \ttfamily arXiv:1805.12028}}.

\bibitem{MiniBooNE:2022emn}
{\normalfont \bfseries MiniBooNE}, A.~A. Aguilar-Arevalo {\em et al.}, ``{\em {MiniBooNE and MicroBooNE Combined Fit to a 3+1 Sterile Neutrino Scenario}},'' \href{http://dx.doi.org/10.1103/PhysRevLett.129.201801}{Phys. Rev. Lett. {\normalfont \bfseries 129} (2022) no.~20, 201801}, \href{http://arxiv.org/abs/2201.01724}{{\normalfont \ttfamily arXiv:2201.01724}}.

\bibitem{Stockdale:1984cg}
I.~E. Stockdale {\em et al.}, ``{\em {Limits on Muon-Neutrino Oscillations in the Mass Range $30<\Delta m^2 < 1000\:\mathrm{eV}^2/c^4$}},'' \href{http://dx.doi.org/10.1103/PhysRevLett.52.1384}{Phys. Rev. Lett. {\normalfont \bfseries 52} (1984)  1384}.

\bibitem{Dydak:1983zq}
F.~Dydak {\em et al.}, ``{\em {A Search for Muon-neutrino Oscillations in the $\Delta m^2$ Range $0.3-90\:\mathrm{eV}^2$}},'' \href{http://dx.doi.org/10.1016/0370-2693(84)90688-9}{Phys. Lett. B {\normalfont \bfseries 134} (1984)  281}.

\bibitem{IceCube:2016rnb}
{\normalfont \bfseries IceCube}, M.~G. Aartsen {\em et al.}, ``{\em {Searches for Sterile Neutrinos with the IceCube Detector}},'' \href{http://dx.doi.org/10.1103/PhysRevLett.117.071801}{Phys. Rev. Lett. {\normalfont \bfseries 117} (2016) no.~7, 071801}, \href{http://arxiv.org/abs/1605.01990}{{\normalfont \ttfamily arXiv:1605.01990}}.

\bibitem{Abbasi:2024ktc}
R.~Abbasi {\em et al.}, ``{\em {Search for a light sterile neutrino with 7.5 years of IceCube DeepCore data}},'' \href{http://arxiv.org/abs/2407.01314}{{\normalfont \ttfamily arXiv:2407.01314}}.

\bibitem{MINOS:2017cae}
{\normalfont \bfseries MINOS+}, P.~Adamson {\em et al.}, ``{\em {Search for sterile neutrinos in MINOS and MINOS+ using a two-detector fit}},'' \href{http://dx.doi.org/10.1103/PhysRevLett.122.091803}{Phys. Rev. Lett. {\normalfont \bfseries 122} (2019) no.~9, 091803}, \href{http://arxiv.org/abs/1710.06488}{{\normalfont \ttfamily arXiv:1710.06488}}.

\bibitem{Super-Kamiokande:2014ndf}
{\normalfont \bfseries Super-Kamiokande}, K.~Abe {\em et al.}, ``{\em {Limits on sterile neutrino mixing using atmospheric neutrinos in Super-Kamiokande}},'' \href{http://dx.doi.org/10.1103/PhysRevD.91.052019}{Phys. Rev. D {\normalfont \bfseries 91} (2015)  052019}, \href{http://arxiv.org/abs/1410.2008}{{\normalfont \ttfamily arXiv:1410.2008}}.

\bibitem{KARMEN:2002zcm}
{\normalfont \bfseries KARMEN}, B.~Armbruster {\em et al.}, ``{\em {Upper limits for neutrino oscillations $\bar\nu_\mu\rightarrow\bar\nu_e$ from muon decay at rest}},'' \href{http://dx.doi.org/10.1103/PhysRevD.65.112001}{Phys. Rev. D {\normalfont \bfseries 65} (2002)  112001}, \href{http://arxiv.org/abs/hep-ex/0203021}{{\normalfont \ttfamily arXiv:hep-ex/0203021}}.

\bibitem{NOMAD:2003mqg}
{\normalfont \bfseries NOMAD}, P.~Astier {\em et al.}, ``{\em {Search for nu(mu) ---\ensuremath{>} nu(e) oscillations in the NOMAD experiment}},'' \href{http://dx.doi.org/10.1016/j.physletb.2003.07.029}{Phys. Lett. B {\normalfont \bfseries 570} (2003)  19--31}, \href{http://arxiv.org/abs/hep-ex/0306037}{{\normalfont \ttfamily arXiv:hep-ex/0306037}}.

\end{thebibliography}\endgroup
\end{document}